%% file: revised.tex
\documentclass[fleqn]{article}
\usepackage{latexsym,url,epsfig}
\usepackage{fullpage}
\usepackage{enumerate}
\usepackage{umlaut}
\usepackage{theorem}
\usepackage{floatflt}
\usepackage{graphicx}
\usepackage{calc}
\usepackage{amssymb,amsmath,bbm}

\newcommand{\marginrem}[1]%
   {%
     \mbox{}\marginpar{\textsl{\fbox{\vbox{\tiny\raggedright\hspace{0pt}#1}}}}%
}%
\newcommand{\textrem}[1]%
   {%
     \mbox{}\marginpar{\textsl{\fbox{\vbox{\tiny\raggedright\hspace{0pt}#1}}}}%
}%
\def\squareforqed{\hbox{\rlap{$\sqcap$}$\sqcup$}}
\def\myqed{\ifmmode\squareforqed\else{\unskip\nobreak\hfil
\penalty50\hskip1em\null\nobreak\hfil\squareforqed
\parfillskip=0pt\finalhyphendemerits=0\endgraf}\fi}
\theoremstyle{break}
\newtheorem{theo}{Theorem}
\newtheorem{defi}[theo]{Definition}
\newtheorem{prop}[theo]{Proposition}

\newcommand{\old}[1]{{}}
\newcommand{\uncomment}[1]{{}}

\newcommand{\N}{{\mathbbm{N}}}
\newcommand{\R}{{\mathbbm{R}}}
\newcommand{\proof}{{\bf Proof. }}
\newcommand{\qed}{{\hfill$\Box$}}
\input{epsf}

\begin{document}

\title{Online Searching with Turn Cost}

\author{
Erik D.\ Demaine\\
Computer Science and Artificial Intelligence Laboratory\\
MIT\\
Cambridge MA, USA\\
{\tt edemaine@mit.edu}
\and
S\'{a}ndor P.\ Fekete\\
Department of Mathematical Optimization\\
Braunschweig University of Technology\\
Braunschweig, Germany\\
{\tt sandor.fekete@tu-bs.de}
\and
Shmuel Gal\\
Department of Statistics\\
University of Haifa\\
Haifa, Israel\\
{\tt sgal@univ.haifa.ac.il}
}

\date{}
\maketitle

\begin{abstract}
We consider the problem of searching for an object on a line
at an unknown distance OPT from the original
position of the searcher, in the presence of a cost of $d$
for each time the searcher changes direction.
This is a generalization of the well-studied linear-search problem.
We describe a strategy that is guaranteed to find the object
at a cost of at most $9 \cdot \mathrm{OPT} + 2d$,
which has the optimal competitive ratio~$9$ with respect to 
OPT plus the minimum corresponding additive term.
Our argument for upper and lower bound
uses an infinite linear program, which we solve by
experimental solution of an infinite series of
approximating finite linear programs,
estimating the limits, and solving the resulting recurrences
for an explicit proof of optimality.
We feel that this technique is interesting in its own right
and should help solve other searching problems.
In particular, we consider the {\em star search} or {\em cow-path problem}
with turn cost,
where the hidden object is placed on one of $m$ rays emanating
from the original position of the searcher.
For this problem we give a tight bound of
$\left(1+2\frac{m^m}{(m-1)^{m-1}}\right) \mathrm{OPT}
+ m \left(\left(\frac{m}{m-1}\right)^{m-1}-1\right) d$.
We also discuss tradeoff between the corresponding coefficients and
we consider randomized strategies on the line.
\end{abstract}




\section{Introduction}
\label{sec:intro}

\subsection{Search games}
Searching for an object is one of the fundamental issues of everyday life, and 
also one of the basic algorithmic problems that need to be mastered in the
context of computing \cite{K73}. If the object is located in a bounded domain
(say, in one of $n$ discrete locations), then the worst-case complexity is 
obvious: In accordance with our everyday experience, an imaginary ``hider''
may have placed the object in the very last place where we look
for it.

More challenging is the scenario of searching in an unbounded domain. 
The classic prototype is the {\em linear-search problem},
which was first proposed by Bellman \cite{B63} and, independently,
by Beck\cite{B64}: 
An (immobile) object is located on the real line according to a known
probability distribution. A searcher, whose maximum velocity is one, starts
from the origin $O$ and wishes to discover the object in minimum expected
time. It is assumed that the searcher can change the direction of her motion
without any loss of time. It is also assumed that the searcher cannot see
the object until she actually reaches the point at which the object is
located; the time elapsed until this moment is the cost function. 
Originally, the problem was presented in a Bayesian context, assuming that
the location of the object is given by a known probability distribution $F$,
but what can we do if we do not know $F$? This situation is quite common and a
natural approach for dealing with it is to try to find search trajectories
that will be effective against all possible distributions. Can such
``universal'' trajectories be found?

For this purpose, it is useful to consider a game between a ``searcher''
S and a ``hider'' H. As the time necessary for the searcher to locate
the object may be arbitrarily high (as the object may be hidden far
from the origin), a useful measure for the performance of a search strategy
is the {\em competitive ratio}: This is the supremum of the ratio
between the time the searcher actually travels and the time she would have 
taken if she had known the hiding place. The competitive ratio is 
a standard notion in the context of online algorithms; see \cite{FW98,GKR01}
for recent overviews. As the supremum has to be taken over all possible
events of a game (or all possible sequences of events in the case
of an online problem), it is quite useful to imagine these events 
chosen by a powerful adversary, who knows the strategy of the searcher.
We will focus on a resulting primal-dual modeling further down.

For the linear-search problem, the optimal competitive ratio is 9,
as was first shown by Beck and Newman \cite{BN70}: The searcher should alternate
between going to the right and to the left, at each iteration doubling 
her step size. By placing the object at one of the points just beyond 
a turning point of the searcher, the hider can actually assure that this
ratio of 9 is best possible.

The linear-search problem has been rediscovered, re-solved, and generalized 
independently by a number of researchers. One such generalization is
the {\em star search} (first solved by Gal \cite{G74}), 
where the searcher
has to locate the object on one of $m$ rays emanating from the origin;
thus, the linear-search problem is a special case for $m=2$.
See \cite{BCR88,BCR93} for a rediscovery and some extensions, as well as
\cite{K93}. More recent results and references can be found 
in \cite{LS98}, which computes the optimal
solution including lower-order terms as a function of the
distance OPT to the object, which need not be known
to the robot to obtain optimal behavior including these terms.

\subsection{Geometric trajectories and turn cost}
One common feature of optimal trajectories for the linear-search
problem and variants is that the step size is a geometric
{sequence.}
 For the star search, the step size increases by a constant factor 
of $\frac{m}{m-1}$ at each iteration, and the overall competitive
factor works out to be $\left(1+2\frac{m^m}{(m-1)^{m-1}}\right)$.
Furthermore, it can be shown that under certain assumptions,
{\em any} unbounded optimal search trajectory has to be a geometric 
sequence \cite{G72,GC76}; see \cite{AG03} for details and citations.

\begin{figure}[htbp]
\begin{center}
\epsfig{figure=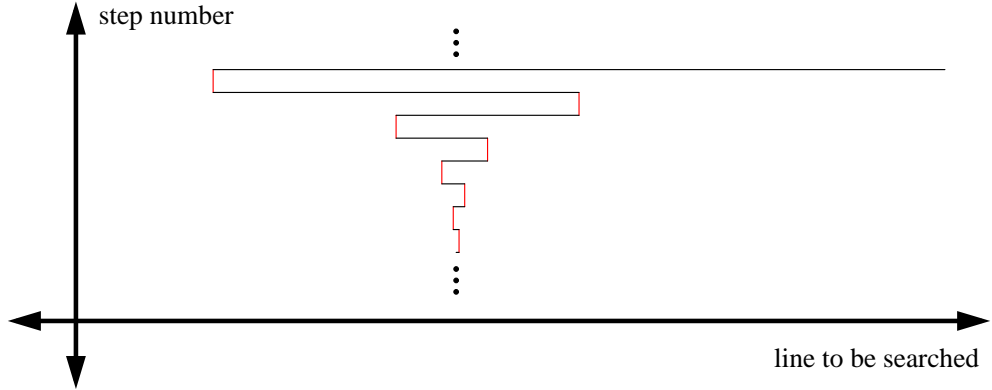,width=0.8\linewidth}
\end{center}
\caption{\label{fig:wiggle}
  An optimal search trajectory for the linear-search problem:
  A geometric sequence without a first step.}
\end{figure}

While these geometric trajectories are quite elegant from a mathematical
point of view, there is a serious downside: As they ``start'' with an 
infinite sequence of infinitesimal steps, they are neither practical,
nor is the necessary time realistic. (See Figure~\ref{fig:wiggle}.)
So far, the issue of the infinitesimal startup has been avoided,
either implicitly or explicitly; e.g. \cite{LS98} says
``In order to avoid this problem we assume that a lower bound of one for 
the distance to the target $t$ is known.''
It should be noted that \cite{LS98,JS01}
have dealt with the ``upper'' part of the infinite sequence
by assuming that an upper bound on the
distance to the object is known in advance.

In this paper, we study a clean and simple way to avoid the problems
of geometric sequences without a first step, by assuming a constant
turn cost $d$ for changing direction. This assumption
is natural and realistic, as any reasonable scenario
incurs some such cost for turning. We describe optimal trajectories
for this scenario; as it turns out, they are generalizations
of the optimal trajectories for the linear-search problem without turn
cost, and have the same asymptotic behavior as $d\rightarrow 0$.
The only previous work we are aware of that mentions turn cost
in the context of the linear-search problem is \cite{l-sbud-96},
which considers (Section 2.5, p.25f.) a kinetic model, where turning
takes a certain amount of time for braking and 
{accelerating.}
It is shown that assuming a lower bound on OPT, a competitive
ratio of 9 is still best possible for large OPT. (``Thus, as the
braking time is of no 
{relevance}
for large $n$, we know that the competitive
ratio must be at least 9 as well.'')
We go beyond this observation by cleanly quantifying the impact of
turn cost on the overall search cost, and doing away with the assumption
of a lower bound on OPT.

\subsection{Other Related Work}
Linear search problems occur in various contexts.
See \cite{AB00} for a study of rendezvous search on a line,
where the objective of two players is to meet as fast as possible;
this turns out to be a double linear-search problem.
\cite{KRT96} studies randomized strategies for the star search
(which is also known as the ``cow-path problem'', motivated by a 
cow searching for the nearest pasture.)
See \cite{K93} for more on the star search, and \cite{HNS99}
for parallel searching. 

Various types of search problems have been considered in a geometric
(mostly two-dimensional) context. Here we only mention
\cite{BS99,HIKL99,LS97,S01}, and the remarkable paper
\cite{HIKK02} that shows that online searching in a simple polygon
can be performed with a competitive ratio of not more than 26.5.
A very recent application in the context of robotics is described in
\cite{fkn-osar-04,fkn-sar-04}, where a robot has to stop 
every time it takes a scan of its environment; just like in this
paper, the overall objective 
is to minimize total time until the discovery of an object hidden 
behind a corner, and this time is the sum of travel time and a cost 
for special points.  However, the resulting trajectories
and mathematical tools turns out to be quite different from what is presented
here.

A good overview on search games can be found in the book \cite{G80},
and the more recent book \cite{AG03}.

Relatively little work has been done on geometric optimization
problems with turn cost. The interested reader may find some 
discussion in the paper \cite{ABDFMS01}.

{ 
\subsection{Duality for Linear Programming}
A standard tool for computing the values of two-player games is
linear programming. In fact, studying such games was one of the origins
of linear programming: An optimal strategy for one player can be interpreted
as an optimal primal solution, while an optimal strategy for the 
second player corresponds to an optimal dual solution.
(The reader unfamiliar with the basics of linear programming
may turn to \cite{Chv} for a good introduction.)

For finite games, the following results are well known
(as weak duality and complementary slackness)
and elementary to prove:

\begin{prop}[Weak duality, strong duality, complementary slackness]
\label{prop:linear}
Let $A$ be a finite real matrix and let $b,c$ be vectors of 
appropriate dimensions.
Let $\max c^tx, Ax\leq b, x\geq 0$ be a primal linear program $(P)$, and let
$\min b^t y, y^t A\geq c, y\geq 0$ be the corresponding
dual linear program $(D)$. 

Then {\em weak duality} holds:
For any pair of feasible solutions, $x'$ for $(P)$
and $y'$ for $(D)$, we have $c^tx'\leq b^ty'$;
if $c^tx'= b^ty'$ for any pair of feasible solutions,
then $x'$ is optimal for $(P)$ and $y'$ is optimal 
for $(D)$. 

Furthermore, {\em strong duality} holds:
If $(P)$ is feasible and finitely solvable by $x^*$,
then there exists a dual feasible solution $y^*$
with $c^tx^*= b^ty^*$.

For any such pair of optimal solutions,
{\em complementary slackness} holds:
If a dual variable $y_j$ is positive, then the corresponding
primal constraint must hold with equality;
if a primal constraint does not hold with equality,
then the corresponding dual variable must be zero.
Conversely, any pair of solutions satisfying complementary
slackness is optimal.
\end{prop}
} 

The paper \cite{JS01} also uses linear programs for the analysis of
the cow-path problem; in particular, it uses these tools for
analyzing the scenario where there is a known limit on the distance
to the object. Considering turn cost makes our problem different;
moreover, we use a different perspective of dealing with the issue 
of infinitely many constraints by considering
dual variables for establishing a lower bound.

\subsection{Our Results}
{ 
Motivated by the linear-search problem with turn cost, we achieve a number of results.

$\bullet$ We establish duality results for certain types of
infinite linear programs. As it turns out, this
implies a verification method for the optimality of strategies.
}    

$\bullet$ We show that the linear-search problem in the presence
of turn cost can be characterized by an infinite linear program.

$\bullet$ Using CPLEX, we perform a computational study on the 
sequence of linear programs obtained by relaxing the infinite 
linear program to a linear subproblem with a finite number of constraints. 

$\bullet$ From the computational results, we derive an analytic proof
of the optimal strategies of searcher and hider. As a consequence of duality,
the resulting bounds are tight: an optimal strategy requires $9$OPT$+2d$,
and the optimal search strategy has step size 
$x_i=d\left(2^{i}-1\right)/2$.

$\bullet$ We also consider how the results vary if we allow 
the competitive ratio to increase at the benefit of decreasing the additive term.

$\bullet$ We generalize the above results to the scenario of star search
by showing that the searcher can guarantee finding a solution
within time 
\[\left(1+2\frac{m^m}{(m-1)^{m-1}}\right)\mbox{OPT} 
+ m \left(\left(\frac{m}{m-1}\right)^{m-1}-1\right) d\]
by choosing the strategy 
\[x_i=d\left(\left(\frac{m}{m-1}\right)^{i+1}-1\right)/2.\]
$\bullet$ We show that for randomized strategies for searching on the line
in the presence of turn
cost, the same optimal competitive ratio $q=4.591121\ldots$ can be
achieved as in the scenario without turn cost.

The rest of this paper is organized as follows. 
{
In Section~\ref{sec:inf}, we describe basic results
on infinite linear programs.
} 
In Subsection~\ref{sec:first} we discuss the start
of the search and show that the presence of turn cost always forces
the existence of a first step with step length bounded from below. 
In Subsection~\ref{sec:lp} we derive an infinite linear program for the
value of the game. Subsection~\ref{sec:cplex} describes the results of a
computational study; a clean analysis, with a mathematical proof of 
optimality of the derived strategies, is given in Section~\ref{sec:proof}.
Subsection~\ref{sec:tradeoff} discusses the tradeoff between the coefficients
of OPT and $d$.
Section~\ref{sec:star} describes the extension to star search.
Section~\ref{sec:rand} gives a brief discussion on randomized
strategies on the line.
Some concluding thoughts are presented in Section~\ref{sec:conc}.

\section{Infinite Linear Programs}
\label{sec:inf}
At first glance, linear programming and duality are not easily applicable
to the search games described above, as the games are unbounded.
However, we demonstrate that it makes sense to consider
{\em infinite linear programs}. As it turns out, we can still construct
primal and dual solutions, and prove optimality by applying weak duality.
For this purpose, we construct infinite linear programs as the limit
of increasingly large finite linear programs, obtained by 
successively adding variables and constraints. In the limit, a
solution is an infinite sequence, instead of a finite vector;
the scalar product of two vectors turns into the series
of component products.
When studying convergence, it helps to think of each solution
vector of a finite linear program as a sequence that has zeroes
for all unused variables.

\begin{defi}[Infinite linear programs]
\label{def:inflp}
Let $c=(c_i)_{i\in \N}$ and $b=(b_j)_{j\in \N}$ be sequences of real
numbers, and let $A=(a_{ij})_{i,j\in\N}$ be a doubly indexed sequence
of real numbers.  Analogously, let 
$x=(x_i)_{i\in \N}$ and $y=(y_j)_{j\in \N}$ be sequences of real
variables.
Denote by $c^{(k)},b^{(k)}, x^{(k)}, y^{(k)}$
the vectors $(c_1\ldots,c_k)$, $(b_1\ldots,c_k)$,
$(x_1\ldots,x_k)$, and $(y_1\ldots,y_k)$, respectively,
and by $A^{(k)}$ the matrix $(a_{ij})_{i,j\in\{1,\ldots,k\}}$.
Furthermore, we use the symbolic notation
$c^t x:= \sum_{i=1}^{\infty}c_ix_i$ for the scalar product of two
sequences.
Let $(P^{(k)})$ be the $k$th subprogram, given by
$\max {c^{(k)}}^tx^{(k)}, A^{(k)}x^{(k)}\leq b^{(k)},x^{(k)}\geq 0$, and denote 
by $(D^{(k)})$ the corresponding $k$th dual linear program,
given by
$\min {b^{(k)}}^t y^{(k)}, {y^{(k)}}^t A^{(k)}\geq c^{(k)}, y^{(k)}\geq 0$.
We say a sequence $x=(x_i)_{i\in\N}$ of nonnegative real numbers
is a {\em feasible solution} for the infinite linear program
given by $A,b,c$, if, for all $j\in\N$, the constraint
$\sum_{i=1}^{\infty}a_{ij}x_i\leq b_j$ holds; i.e.,
for all $k\in\N$, $x^{(k)}$ is feasible for $(P_k)$. 
An {\em infinite linear program} $(P)$, denoted by 
$\max c^tx, Ax\leq b, x\geq 0$,
is the problem of finding a feasible solution that
maximizes $c^t x$.
A sequence $y=(y_j)_{j\in\N}$ of nonnegative real numbers
is an {\em infinite dual solution}, if for all $i$, 
$c_i\leq\sum_{j=1}^{\infty}A_{ij}y_j<\infty$.
\end{defi}

The following provides a way of establishing bounds for and optimality
of feasible solutions for infinite linear programs.

\bigskip
\begin{theo}[Weak duality for infinite LPs]
\label{th:weakdual}
Let $\max {c}^tx, Ax\leq b,x\geq 0$ be a an infinite linear program.
Assume that the set $I:=\{i\in\N\mid c_i\neq 0\}$ is finite,
and that for any $i\in\N$, all but finitely many $a_{ij}$ have the same sign.
Let $x$ be a feasible solution for $(P)$,
and let $y$ be an infinite dual solution. 
Then weak duality applies: ${c}^tx\leq{b}^t y$.
Moreover, if ${c}^tx={b}^t y$, then $x$ is optimal. 
\end{theo}

\proof
By the above assumptions, we have
\[c^tx=\sum_{i=1}^{\infty}c_ix_i=\sum_{i\in I}c_ix_i.\]
By assumption on $y$ it follows from $x_i\geq 0$ that
\[c^tx\leq 
\sum_{i\in I}x_i\sum_{j=1}^{\infty}a_{ij}y_j
=
\sum_{i\in I}\sum_{j=1}^{\infty}x_i a_{ij}y_j
\leq
\sum_{i=1}^{\infty}\sum_{j=1}^{\infty}x_i a_{ij}y_j.\]
For any $i$, only finitely many of the $a_{ij}$
have a sign different from the others, so
the converging series $\sum_{j=1}^{\infty}x_i a_{ij}y_j$
is absolutely convergent. Furthermore, only finitely many
$c_i$ can be negative, implying that all all but finitely many
$\sum_{j=1}^{\infty}a_{ij}y_j$ are nonnegative.
Thus, we may swap summations,
getting
\[c^tx\leq 
y^t(Ax)=
\sum_{j=1}^{\infty}\sum_{i=1}^{\infty}x_i a_{ij}y_j
=
\sum_{i=1}^{\infty}\sum_{j=1}^{\infty}x_i a_{ij}y_j
=(y^tA)x
\leq
\sum_{j=1}^{\infty}b_j y_j = b^ty,\]
as claimed.
\qed

\bigskip
The following motivates our approach to solving infinite linear
programs by considering the sequence of finite subproblems.
Assuming convergence of the sequences of primal and dual solutions,
we get strong duality:

\bigskip
\begin{theo}[Strong duality for infinite LPs]
\label{th:strongdual}
Let $\max {c}^tx, Ax\leq b,x\geq 0$ be a an infinite linear program.
Assume that the set $I_+:=\{i\in\N\mid c_i>0$ is finite.
Let for all $k\in \mathbbm{N}$, $(P^{(k)})$ be feasible and bounded,
and the sequence of primal optimal solutions be bounded;
let $x^{*(k)}$ be a sequence of corresponding primal optimal solutions. 
For the corresponding sequence
$\min {b}^t y^{(k)}, {y^{(k)}}^t A^{(k)}\geq c, y^{(k)}\geq 0$ 
of dual linear programs $(D_k)$, 
let $y^{*(k)}$ be a sequence of dual optimal solutions.
Suppose that as $k$ tends to infinity,
$x^{*(k)}$ converges componentwise to $x^*$ and
$y^{*(k)}$ converges componentwise to $y^*$.
Then $x^*$ is an optimal solution for the 
infinite linear program $(P)$.
Moreover, strong duality holds, i.e., ${c^{*}}^tx^*={b^*}^t y^*$. 
\end{theo}

\proof
We start by establishing primal feasibility of $x^*$:
Suppose that inequality $(j)$ was violated by $x^*$,
i.e., $\sum_{i=1}^{\infty}a_{ij}x_i>b_j$.
As $a_{i\ell}=0$ for $\ell>j$,
the first $j$ components of $x^*$ involved in $(j)$
form a point that has a positive distance from the set 
$\{z\in\R^j\mid\sum_{i=1}^j a_{ij}z_i\leq b_j\}$.
As for all $k\geq j$, inequality $(j)$ is part of $(P^{(k)})$,
it must be satisfied by $x^{*(k)}$. This contradicts the fact
that $x^{*(k)}$ converges towards $x^*$.

For $k> \max\{i\in I_+\}$, the sequence of objective values
${c^{(k)}}^tx^{*(k)}$ is monotonically decreasing,
as for $k\leq \ell$, the feasible set for $(P^{(\ell)})$ is a 
subset of the feasible set for $(P^{(k)})$.
As the sequence of objective values is bounded, 
$\lim_{k\rightarrow\infty} {c^{(k)}}^tx^{*(k)}$ exists.
For any $k$, 
we have ${c^{(k)}}^tx^{*(k)}={b^{(k)}}^t y^{(k)}$ by strong duality. 
Thus, we have
\[c^tx^*=\lim_{k\rightarrow\infty} {c^{(k)}}^tx^{*(k)} =
  \lim_{k\rightarrow\infty} {b^{(k)}}^t y^{*(k)}=b^ty^*.\]
By Theorem~\ref{th:weakdual}, this implies optimality of $x^*$.
\qed

\bigskip
Finally, complementary slackness holds; this will turn out to be
a useful tool for determining dual variables, and thus for proving optimality.

\bigskip
\begin{theo}[Complementary slackness for infinite LPs]
\label{th:compslack}
Let $\max {c}^tx, Ax\leq b,x\geq 0$ be a an infinite linear program.
Assume that the set $I:=\{i\in\N\mid c_i\neq 0\}$ is finite,
and that for any $i\in\N$, all but finitely many $a_{ij}$ have the same sign.
Let $x$ be a feasible solution for $(P)$, and 
$y$ be an infinite dual solution.
Then $x$ and $y$ are optimal if and only if
the following complementary slackness conditions hold:

(a) For any $i\in\N$ with $\sum_{j=1}^{\infty}y_ja_{ij}>c_i$, $x_i=0$ holds.

(b) For any $i\in\N$ with $x_i>0$, $\sum_{j=1}^{\infty}y_ja_{ij}=c_i$ holds.

(c) For any $j\in\N$ with $\sum_{i=1}^{\infty}a_{ij}x_i<b_i$, $y_j=0$ holds.

(d) For any $j\in\N$ with $y_j>0$, $\sum_{i=1}^{\infty}a_{ij}x_i=b_i$ holds.
\end{theo}

\proof
By assumptions, all involved series are absolutely convergent. 
Thus we get $(y^tA)x\geq c^tx\Leftrightarrow (y^tA-c)x\geq 0$,
and $y^tb\geq y^t(Ax)\Leftrightarrow y^t(b-Ax)\geq 0$.
Now it is easy to check that $y^tb=c^tx$, iff
the stated conditions hold.
\qed

\bigskip
We believe that the above tools are useful and applicable in 
various game-theoretic
scenarios, even when there is no clear idea of a possible optimal strategy.
We demonstrate the practical applicability
by giving the results of a computational study
performed with CPLEX
\cite{cplex}, 
a commercially available software package for solving
linear programs, and showing how these results enable us to give a tight
(theoretical) analysis of the cost of an optimal search strategy in the 
presence of turn cost. It should be noted that in this paper,
the numerical experiments are only a stepping stone towards 
finding optimal strategies:
The idea is to identify closed-form limiting solutions;
once those are found and verified by using duality, 
numerical accuracy is not an issue. That means that the numerical
results could be omitted without impeding the proof of optimality
of the resulting strategies. However, 
our approach should also prove useful for other problems, 
even if no optimal closed-form
strategy can be deduced from solving finite subsystems: 
In those cases, we can still give bounds on the possible
performance of strategies. Only in those cases, issues of numerical 
stability come into play.

\section{Searching on the Line with Turn Cost}
\subsection{The First Move and an Additive Term}
\label{sec:first}

In the presence of a positive turn cost $d$, the hider can make it
impossible for the searcher to achieve a competitive ratio,
by simply placing the object arbitrarily close to the origin,
on the side that is not picked first by the searcher. Clearly,
this requires a minimum cost of $d$, regardless of OPT.
Moreover, the searcher will be forced to make a second turn
if she starts with a too small (or infinitesimal) first step.
This increases the minimum cost.

We show that the optimal competitive ratio, $c$, remains the same 
even if we add turn cost. Thus, the worst-case time to reach the target is $c\cdot$OPT plus some additive term, denoted $B$, that we wish to minimize.
Determining the minimum value of $B$ is one of the main objectives of this paper.

It is clear that even in the presence of a fixed additive term, 
the searcher will not be able to achieve any competitive
ratio at all if she uses a large number of steps before first
reaching a distance of $d$ from the origin. In particular,
it can easily be seen that she is forced to make a first step of
length $x_1=\Omega(d)$. We will use this observation in the following subsections
for a more careful analysis.

\subsection{An Infinite Linear Program}
\label{sec:lp}

Suppose that the searcher carries out a sequence
of step lengths $x_1,x_2,\ldots$ from the origin, where
$x_1,x_3,\ldots$ are increasing distances to the right,
while $x_2,x_4,\ldots$ are increasing distances to the left.
In the following, we denote these turn positions by 
$p_i=(-1)^{i+1}x_i$.

For a given sequence of step lengths $x_i$, the hider can choose
the possible set of locations $y_i=(-1)^{i+1}(x_i+\varepsilon)$
for an arbitrarily small $\varepsilon$. If the object is placed
at $y_n$, the searcher will only encounter it after traveling
a distance of $(\sum_{i=1}^{n+1} 2x_i ) + x_n + \varepsilon$,
and making $n+1$ turns. Note that for arbitrarily small $d$,
this approaches the linear-search problem, with a competitive ratio
of 9. In order to guarantee this competitive ratio, and
an additive cost of $B$, 
the corresponding search trajectories satisfy
%
\[2x_1+\ldots+2x_{i-2}+3x_{i-1}+2x_i+id+\varepsilon\leq 9(x_{i-1}+\varepsilon)+B.\label{eq:eins}\]
As all $x_i$ are bounded away from zero, and the above condition must hold
for any $\varepsilon>0$, we conclude that
%
%
\[2x_1 + 2x_2 + \ldots + 2x_{i-2} + 3x_{i-1} + 2x_i + id  \leq  9x_{i-1} + B.\label{eq:zwei}\]
or
%
\[2x_1 + 2x_2 + \ldots + 2x_{i-2} - 6x_{i-1} + 2x_i + id  \leq  B.\label{eq:drei}\]

We thus get the following
infinite linear program:

\begin{equation}
\begin{array}{rcrcrcrcrcrcrcr}
&&\min&B\\
2x_1 &&&&&&&&&&&+& d & \leq & B\\
-6x_1 &+& 2x_2 &&&&&&&&&+& 2d & \leq & B\\
2x_1 &-&  6x_2 &+& 2x_3 &&&&&&&+& 3d & \leq & B\\
2x_1 &+&  2x_2 &-& 6x_3 &+& 2x_4 &&&&&+& 4d & \leq & B\\
\vdots &+&  \vdots &-& \vdots &+& \vdots &&&&&+& \vdots & \leq & B\\
2x_1 &+&  2x_2 &+& \cdots &+& 2x_{i-2} &-& 6x_{i-1} &+& 2x_{i} &+& id & \leq & B\\
\vdots &&  \vdots && \vdots && \vdots &&\vdots&&\vdots&& \vdots & 
\leq & B\\
&&&&&&&&&&x_i&&&\geq & 0.\\
\end{array}
\label{lpi}
\end{equation}

\subsection{Solving the Sequence of Linear Programs}
\label{sec:cplex}

Even with sophisticated software, it is impossible to solve an
infinite linear program to optimality. However, for each
$n$, the first $n$ linear inequalities describe a relaxation
of the overall program. Denote $\lambda=B/d$, and let $\lambda_n$
be the optimal value when only considering the first $n$
constraints. Then any
$\lambda_n$ is a lower bound on the overall $\lambda$.
Furthermore, the sequence of primal optimal values $x_i^{(n)}$
and dual optimal values $y_i^{(n)}$ should converge, if there
is any hope for an overall solution.

We have solved a number of these relaxations by using CPLEX 7.1.
The computational results are shown in Table~\ref{tab:lp}.
Shown are the number of constraints that we used, the optimal
value $\lambda_n$, the first five primal variables, and the 
first five dual variables. To allow numerical computation, $d$
was normalized to 1. As each $\lambda_n$ is a valid lower bound,
numbers were truncated, not rounded.

\begin{table}
\begin{center}
{\scriptsize
\begin{tabular}{|r|r|r|r|r|r|r|r|r|r|r|r|}
\hline
$n$ & $\lambda_n$ & $x_1^{(n)}$ & $x_2^{(n)}$ & $x_3^{(n)}$ & $x_4^{(n)}$ & $x_5^{(n)}$ & $y_1^{(n)}$ & $y_2^{(n)}$ & $y_3^{(n)}$ & $y_4^{(n)}$ & $y_5^{(n)}$\\
\hline
1 &  \ 1.0000  \ &  \ 0.0000  \ &  &   &   &  &  &  &  &  & \\
2 &  \ 1.2500  \ &  \ 0.1250  \ &  \ 0.0000  \ &    \ &    \ &    \ &  \ 0.7500  \ &  \ 0.2500  \ &    \ &    \ & \\
3 &  \ 1.4166  \ &  \ 0.2083  \ &  \ 0.3333  \ &  \ 0.0000  \ &    \ &    \ &  \ 0.6666  \ &  \ 0.2500  \ &  \ 0.0833  \ &    \ & \ \\
4 &  \ 1.5312  \ &  \ 0.2656  \ &  \ 0.5625  \ &  \ 0.6875  \ &  \ 0.0000  \ &    \ &  \ 0.0625 \ &  \ 0.2500  \ &  \ 0.0937 \ &  \ 0.0312  \ & \ \\
5 &  \ 1.6125  \ &  \ 0.3062  \ &  \ 0.7250  \ &  \ 1.1750  \ &  \ 1.3000  \ &  \ 0.0000  \ &  \ 0.6000  \ &  \ 0.2500  \ &  \ 0.1000  \ &  \ 0.0375  \ & \ 0.0125 \ \\
6 &  \ 1.6718  \ &  \ 0.3359  \ &  \ 0.8437  \ &  \ 1.5312  \ &  \ 2.2500  \ &  \ 2.3750  \ &  \ 0.5833  \ &  \ 0.2500  \ &  \ 0.1041  \ &  \ 0.0416  \ &  \ 0.0156 \ \\
7 &  \ 1.7165  \ &  \ 0.3582 \ & \ 0.9330 \ & \ 1.7991 \ & \ 2.9642 \ & \ 4.1607 \ & \ 0.5714 \ & \ 0.2500 \ & \ 0.1071 \ & \ 0.0446 \ & \ 0.0178 \ \\
8 &  \ 1.7509  \ &  \ 0.3754 \ & \ 1.0019 \ & \ 2.0058 \ & \ 3.5156 \ & \ 5.5930 \ & \ 0.5625 \ & \ 0.2500 \ & \ 0.1093 \ & \ 0.0468 \ & \ 0.0195 \ \\
9 &  \ 1.7782  \ &  \ 0.3891 \ & \  1.0563 \ & \ 2.1692 \ & \ 3.9130 \ & \ 6.6284 \ & \ 0.5555 \ & \ 0.2500 \ & \ 0.1111 \ & \ 0.0486 \ & \ 0.0208 \ \\
10 &  \ 1.8001 \ & \ 0.4000 \ & \ 1.1003 \ & \ 2.3011 \ & \ 4.3031 \ & \ 7.5078 \ & \ 0.5500 \ & \ 0.2500 \ & \ 0.1125 \ & \ 0.0500 \ & \ 0.0218 \ \\
20 & \ 1.9000 \ & \ 0.4500 \ & \ 1.3000 \ & \ 2.9000 \ & \ 5.9000 \ & \ 11.5000 \ & \ 0.5250 \ & \ 0.2500 \ & \ 0.1187 \ & \ 0.0562 \ & \ 0.0265 \ \\
30 & \ 1.9333 \ & \ 0.4666 \ & \ 1.3666 \ & \ 3.1000 \ & \ 6.4333 \ & \ 12.8333 \ & \ 0.5166 \ & \ 0.2500 \ & \ 0.1208 \ & \ 0.0583 \ & \ 0.0281 \ \\
40 & \ 1.9500 \ & \ 0.4750 \ & \ 1.4000 \ & \ 3.2000 \ & \ 6.7000 \ & \ 13.5000 \ & \ 0.5125 \ & \ 0.2500 \ & \ 0.1218 \ & \ 0.0593 \ & \ 0.0289 \ \\
50 & \ 1.9600 \ & \ 0.4800 \ & \ 1.4200 \ & \ 3.2600 \ & \ 6.8600 \ & \ 13.9000 \ & \ 0.5100 \ & \ 0.2500 \ & \ 0.1225 \ & \ 0.0600 \ & \ 0.0293 \ \\
100 & \ 1.9800 \ & \ 0.4900 \ & \ 1.4600 \ & \ 3.3800 \ & \ 7.1800 \ & \ 14.7000\ \ & \ 0.5050 \ & \ 0.2500 \ & \ 0.1237 \ & \ 0.0612 \ & \ 0.0303 \ \\
200 & \ 1.9900 \ & \ 0.4950 \ & \ 1.4800 \ & \ 3.4400 \ & \ 7.3400 \ & \ 15.1000 \ & \ 0.5025 \ & \ 0.2500 \ & \ 0.1243 \ & \ 0.0618 \ & \ 0.0307 \ \\
400 & \ 1.9950 \ & \ 0.4975 \ & \ 1.4900 \ & \ 3.4700 \ & \ 7.4200 \ & \ 15.3000 \ & \ 0.5012 \ & \ 0.2500 \ & \ 0.1245 \ & \ 0.0621 \ & \ 0.0310 \ \\
\hline
\end{tabular}
}
\caption{Solutions for a number of linear subsystems.}
\label{tab:lp}
\end{center}
\end{table}

The table illustrates several points:
\begin{enumerate}
\item Even small 
{ 
subsystems
} 
 may have ``ugly'' solutions, indicating
{ 
a } 
relatively fast increasing effort for trying to establish lower
bounds by analyzing subsystems manually.
\item Convergence is rather slow; in fact, 
{ 
it seems to be 
logarithmic, as doubling
the size of the system appears to cut the remaining error in half.} 
This and the tediousness of the solutions
make it rather difficult to give an explicit closed formula for the 
objective values, which could be used for analyzing the limit.
This justifies the use of powerful tools like CPLEX in order to
get good estimates quickly.
\item Actually closing the remaining gap in order to prove that the lower bound
of 9OPT$+2d$ is tight requires considering the whole infinite linear program.
\end{enumerate}

We will demonstrate in the following subsection how the latter point
can be carried out analytically.

\subsection{Provably Optimal Strategies}
\label{sec:proof}

To establish optimal strategies for the infinite linear program,
and thus a pair of optimal strategies, we start by describing a
feasible dual solution that yields a lower bound of 2 for the objective value
$\lambda$.

Consider $y_j=\frac{1}{2^j}$. We will show that with these dual multipliers,
the linear combination of the first $n$ coefficients for  any variable $x_j$ 
tends to $0$, as $n$ approaches infinity. 
More precisely, 
taking a linear combination of all inequalities in (\ref{lpi}), with coefficient
$y_j$ used for inequality $j$, yields

\begin{eqnarray*}
&&   2\left(\sum_{j=1}^{\infty} y_j - 8y_2\right) x_1 +  2\left(\sum_{j=2}^{\infty} y_j - 8y_3\right) x_2 + \cdots 
+ 2\left(\sum_{j=i}^{\infty} y_i - 8y_{i+2}\right) x_{i}
+ \cdots + \left(\sum_{j=1}^{\infty} j y_j\right) d \\
\leq& &\left(\sum_{j=1}^{\infty}y_j\right)  B\label{comb}
\end{eqnarray*}

Using $y_j=\frac{1}{2^j}$, the involved series and coefficients do converge:
The coefficient of $x_i$ becomes
\[- \frac{8}{2^{i+1}} + \sum_{j=i}^{\infty}\frac{2}{2^j} = \frac{1}{2^i}\left(-4 + 2\sum_{i=0}^{\infty} \frac{1}{2^j}\right).\]
Thus, this turns out to be $0$.

Similarly, the coefficient of $d$ becomes $\sum_{i=j}^n \frac{j}{2^j}$,
which tends to $2$ as $n$ grows.

Finally, the coefficient of $B$ on the right hand side of the 
inequality is the geometric series $\sum_{i=1}^n \frac{1}{2^j}$,
which tends to $1$.

Therefore, the derived inequality simplifies to the lower bound
\[2d\leq B.\]

To see that $2d$ is also an upper bound, and thus an optimal
solution for $B$, consider $x_i=(2^i-\frac{1}{2})d$.
For this particular set, the $j$th inequality becomes
\[\left(\sum_{h=1}^{i} 2 x_h\right) - 8 x_{j-1} + jd \leq B,\]
which simplifies to 
\[2d \leq B.\]
Thus, the linear program becomes
\begin{center}
\begin{equation}
\begin{array}{rcc}
&\min&\lambda\\
2 & \leq & \lambda.
\end{array}
\end{equation}
\end{center}

Trivially, this is solved by $\lambda=2$,
implying that the given $x_i$ and $\lambda=2$ do 
indeed provide a feasible solution with objective value 2.
By duality of linear programming, this is optimal.

It should be noted that the primal and dual solutions satisfy 
complementary slackness, as all constraints hold with equality.
Note that our solution to the linear-search problem with turn cost achieves a 
competitive ratio of 9,
which must be the optimal value because the time to find the target increases
with $d>0$.

We summarize:

\begin{theo}
\label{th:linear}
In the presence of turn cost $d$ for the linear-search problem, 
the searcher can guarantee a solution within time 9OPT$+2d$
by choosing the search strategy $x_i=d(2^i- 1)/2$.
The additive term $2d$ is minimal subject to the optimal competitive ratio, 9.
\end{theo}

Note that the overall total cost spent on turning is about $\log_2$ OPT.

\subsection{Tradeoff between Coefficients}
\label{sec:tradeoff}
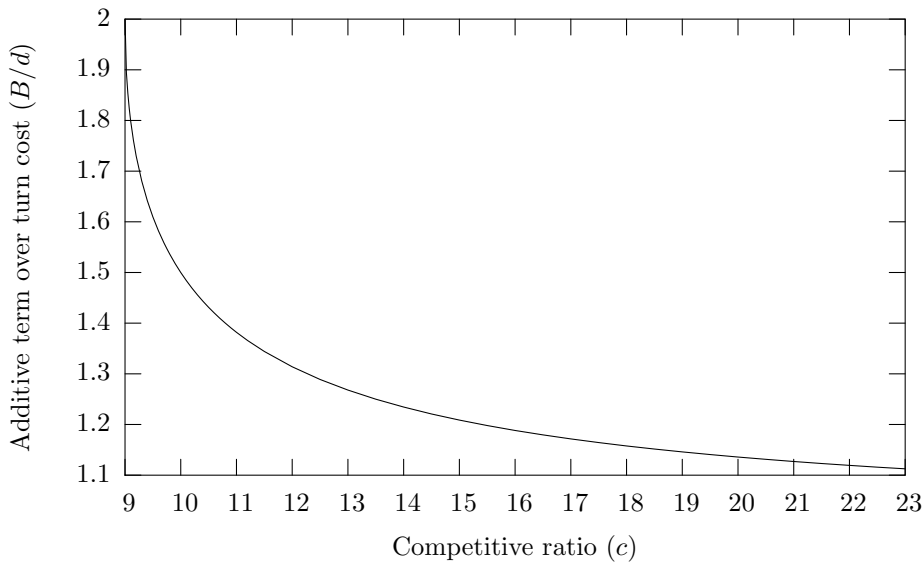
\begin{figure}[htbp]
\begin{center}
\input{sandor.tex}
\end{center}
\caption{\label{fi:tradeoff}
  Tradeoff between competitive factor and turn cost.}
\end{figure}

The term $2d$ is best possible if we want 
{to}
maintain the best possible competitive
factor of $9$. It may be desirable to improve the former term, while
allowing an increase in the latter. For any bound $c\geq 9$
on the competitive ratio, the best possible $B$ can be computed
by using our above approach: In the system (\ref{lpi}), replace
all coefficients $-6$ by $3-c$. Conversely, we can compute
the best possible $c$ for any $B/d\in (1,2]$.
The resulting tradeoff curve is shown in Figure~\ref{fi:tradeoff}.
This curve was obtained experimentally using
CPLEX.  We have not yet characterized it analytically, though we expect
that to be possible.
It is not hard to see that $1$ is a lower bound for any $B/d$, and
that the best possible $B/d$ tends to $1$ for $c$ approaching infinity.

\section{Star Search}
\label{sec:star}

For the problem with no turn cost, Gal~\cite{G80}
proved that the sequence of turning points corresponding to an optimal
search trajectory has to be cyclic.
In the following, we show how to generalize the above results to
the problem of star search in the presence of turn cost.

As before, we consider a sequence $x_1,x_2,\ldots$ of steps
that cycle through the $m$ rays. Suppose we have a turn cost
of $d_1$ on a ray, and $d_2$ at the origin; set $d=d_1+d_2$.
By picking a hiding spot just beyond one of the turning points,
the hider can force the searcher to find the object only
after making moves $x_n,\ldots,x_{n+m-1}$ and returning to the
ray of $x_n$, making a total of $(n+m-1)$ turns.
This takes a time of
$\left(2\sum_{i=1}^{n+m-1}x_i\right) + (n+m-1)d + x_n$
instead of the optimal $x_n$. Without the presence of turn cost,
it is well known that the optimal competitive ratio is
$\left(1+2\frac{m^m}{(m-1)^{m-1}}\right)=:1+2M$.
When the hiding point is close to the start, the searcher
may only get to it when entering the last ray, so we get
the condition
\[\left(2\sum_{i=1}^{m-1}x_i\right) + (m-1)d \leq B. \hfill (m-1)\]
If the hiding point is just beyond the $n$th turning point, we get
the condition
\[\left(2\sum_{i=1}^{n+m-1}x_i\right) + (n+m-1)d + x_n \leq B + 
(1+2M)x_{n},\]
or 
\[\left(2\sum_{i=1}^{n+m-1}x_i\right) + (n+m-1)d \leq B + 
2Mx_{n}.\hfill (n)\]
for the additive term $\lambda$, if it exists.
For convenience, we refer to 
the dual variables corresponding to constraints $(m-1),\ldots,(n),\ldots$
as $y_{m-1},\ldots,y_n,\ldots$. 

Again, finding a strategy $x_1,x_2,\ldots$
that minimizes $B$ subject to the above constraints
can be described as an infinite linear program.
For $m=3,4,5,6$ and various $n$ up to 1000, we solved 
the finite subprograms by using CPLEX. 
Despite some numerical
difficulties, we were able to extrapolate the limits
of the series by making use of the logarithmic convergence.
We obtained the following solutions:
\[x_i = d\left(\left(\frac{m}{m-1}\right)^i-1\right)/2\]
and
\[B = m\left(\left(\frac{m}{m-1}\right)^{m-1}-1\right)d=(M-m)d.\]

Note that finding $x_i$ is also possible without
deriving a closed-form solution directly from CPLEX experiments:
If $y_j>0$ for all $j\in\N$, complementary slackness
(Theorem~\ref{th:compslack}) requires that all
constraints hold with equality.
Then subtracting constraint $(n+m-1)$ from constraint
$(n+m)$ yields the recursion
$2x_{m-1+n}+d=2Mx_n$, which is satisfied
by the above solution.

\begin{theo}
\label{th:star}
In the presence of turn cost $d$ for the star search problem on $m$ rays, 
the searcher can guarantee a solution within time 
\[\left(1+2\frac{m^m}{(m-1)^{m-1}}\right){\rm OPT} 
+ m \left(\left(\frac{m}{m-1}\right)^{m-1}-1\right) d\]
by choosing the search strategy 
\[x_i=d/2\left(\left(\frac{m}{m-1}\right)^{i+1}-1\right).\]
This strategy is optimal.
\end{theo}

\proof
We show that this primal strategy and $B$ satisfy all
constraints of the linear program with equality:
The left-hand side of inequality $(n)$ is
\[2\left(\sum_{i=1}^{n+m-1} x_i\right) + (n+m-1)d.\]
Using $q:=\frac{m}{m-1}$ 
and substituting the values $x_i=d(q^{i}-1)/2$, this simplifies to 
\[\frac{q^{n+m}-q}{q-1}d.\]
On the other hand, by substituting the above values, the right-hand side 
\[B + \left(2\frac{m^m}{(m-1)^{m-1}}\right) x_{n}\]
becomes
\[m\left(q^{m-1}-1\right)d+mq^{m-1}\left(q^n-1\right)d\]
or (using $m=\frac{q}{q-1}$)
\[\frac{q^{n+m}-q}{q-1}d.\]

Therefore, with $B = m\left(\left(\frac{m}{m-1}\right)^{m-1}-1\right)d$,
all constraints are satisfied with equality, regardless of $n$.

It remains to be shown that the above solution is
best possible. In principle, this can be done by
figuring out explicit closed-form expressions
for the dual variables, and using them to verify
optimality. (The interested reader may try this
for $m=3$, where the dual variables turn out to
be $y_j=\frac{4(2^{j-1}+(-1)^{j})}{3^{j+1}}$ for $j\geq m-1=2$,
i.e., the sequence 4/9, 4/27, 4/27, 20/243, 44/729,...)
However, this explicit approach appears to 
be extremely tedious for any $m>3$, and hopeless
for general $m$.
As we will see in the following, 
finding an explicit closed-form expression for dual
variables is not necessary for a proof of optimality.
Instead, we use complementary slackness to derive
a recursive characterization of $y$, and verify
that it satisfies all required conditions.

As any solution $x_1,\ldots$ describing a valid strategy must satisfy $x_i>0$
for all $i\geq 1$, we conclude by Theorem \ref{th:compslack}
that all corresponding dual constraints
must hold with equality; as $c_i=0$ for all $x_i$, 
this means that 
\[\left(2\sum_{j=i}^{\infty}y_j\right) = 2M y_{i+m-1} \hfill (i)\]
must hold for all $i\in\N$,
with $y_1=\ldots=y_{m-2}=0$ for ease of notation.
Noting that we are trying for a nonnegative $y$ 
(and thus an absolutely convergent series $\sum_{j=i}^{\infty}y_j$), 
subtracting this condition $(i)$ for $i=n\geq m-1$ from the condition for 
$i=n+1$ yields
\[y_{n+1} = M (y_{n+m}-y_{n+m-1}),\]
i.e., the recursion
\[y_{n+m} = y_{n+m-1}-\frac{1}{M}y_{n}.\]
 Because the cost coefficient $c_B$ of $B$ is $-1$, we
get the requirement $\sum_{j=m-1}^{\infty}y_j=1$,
which implies $y_m=\ldots=y_{2m-2}=1/M$
by conditions $(1),\ldots,(m-1)$.
Choosing $y_{m-1}=m/M$, we get a well-defined
sequence $y$. Using the initial condition
$y_{2m-2}=1/M$, the recursive condition implies
$y_{n+m}=\frac{1}{M}(1-\sum_{j=m-1}^n y_j)$.
Because of $\frac{m}{M}=\frac{(m-1)^{m-1}}{m^{m-1}}=
\left(1-\frac{1}{m}\right)^m\in [1/3,1/2]$, the series 
$\sum_{j=m-1}^{\infty} y_j=\lim_{n\rightarrow\infty}\sum_{j=m-1}^n y_j$ 
does indeed tend to 1.
(In fact, not only does $y_{j+m}/y_{j}$ tend to $m/M$ for large $j$, 
but the ratio $y_j/y_{j-1}$ tends to $\left(\frac{m-1}{m}\right)$.)
For the given starting values, the sequence $\sum_{j=m-1}^n y_j$
remains below 1, so the sequence $y_{n+m}$ remains nonnegative.

%
%
%
\bigskip
It remains to be shown that $y^tb=c^tx$, 
i.e., $\sum_{j=m-1}^{\infty} j y_j = M-m$.
This follows from 
\begin{eqnarray*}
\sum_{j=m-1}^{\infty}jy_j&=&\sum_{j=m-1}^{2m-2}jy_j+\sum_{j=2m-1}^{\infty}jy_j\\
&=&\sum_{j=m-1}^{2m-2}jy_j+\sum_{j=m-1}^{\infty}(j+m)y_{j+m}\\
&=&\sum_{j=m-1}^{2m-2}jy_j+\sum_{j=m-1}^{\infty}(j+m)(y_{j+m-1}-\frac{1}{M}y_j)\\
&=&(2m-2)y_{2m-2}+\sum_{j=m-1}^{2m-3}jy_j+\sum_{j=m-1}^{\infty}(j+m-1)y_{j+m-1} +\sum_{j=m-1}^{\infty}y_{j+m-1}\\
&& -\sum_{j=m-1}^{\infty}\frac{1}{M}jy_j -\sum_{j=m-1}^{\infty}\frac{m}{M}y_j\\
&=&\frac{2m-2}{M}+\sum_{j=m-1}^{\infty}jy_{j}+\left(1-\sum_{j=m-1}^{2m-3}y_{j}\right) -\sum_{j=m-1}^{\infty}\frac{1}{M}jy_j-\frac{m}{M},
\end{eqnarray*}
hence
\[\sum_{j=m-1}^{\infty}jy_j=2m-2+(M-m-(m-2))-m=M-m,\]
as claimed.
\qed

\section{Randomized Strategies}
\label{sec:rand}

Consider the search on the line with turn cost.
It is natural to consider randomized strategies for both players:
The particular choice of search strategy or hiding position
depends on the outcome of a random event that is not known
in advance. It is known that the optimal competitive ratio $q=1+a$ for
online searching on the line {\em without} turn cost is
given by the solution of the equation satisfying
$\frac{a+1}{\ln a}=a$, or $q=4.591121\ldots$; see \cite{AG03},
pp.\ 129/130. Clearly, the optimal coefficient $q_t$ of OPT in the
presence of turn cost must satisfy  $q_t\geq q$. To see that there is
indeed a strategy that achieves $q_t=q$, consider a modified scenario
without turn cost as follows:

\begin{enumerate}
\item The searcher only locates the hider after passing the hider
by a length of at least $d/2$.
\item When discovering the hider, the cost is the distance traveled,
minus $d/2$.
\end{enumerate}

Then the resulting calculations from \cite{AG03} translate
precisely to the scenario with turn cost. Using the same turning
points $x_i\geq d/2$, we get a mixed strategy that assures a
cost of $q(OPT+d/2) -d/2 = q OPT + d(q-1)/2$.
The additive constant can be improved, as the terms below $d/2$ are not
used. We conclude:
\begin{theo}
\label{th:rand}
For randomized strategies in the presence of turn cost, the optimal
coefficient of OPT is the same as without turn cost.
\end{theo}
Again it is possible to consider the optimal coefficient of the turn-cost
factor, and also consider the tradeoff between both coefficients.
We leave this to future research.

\section{Conclusions}
\label{sec:conc}

In this paper we have considered the linear-search problem in the presence
of turn cost. We have shown that this extends the well-studied case
without turn cost, and established a performance guarantee of 9OPT$+2d$.
We also extended or results to the general star search on $m$ rays
(also known as the cow-path problem), and showed that this problem
can also be resolved by using an infinite sequence of linear programs.

We believe that our methods and results can be easily extended to various
other problems that have been studied; in particular, it should not be too 
hard to give explicit estimates for the lower-order terms, which show up
if the distance OPT to the hidden object is known to be bounded by some $D$:
This only requires giving an explicit estimate for the solutions
of subsystems of size $n=\Theta(\log D)$.

Just like the cow-path problem extends to various geometric
scenarios, we expect that there are also many other problems for which
the cost of changing the search direction plays an important role.

\section*{Acknowledgments}

We thank two anonymous referees for helpful suggestions that helped
to extend the scope of this paper and improve overall presentation.
Parts of this research were supported by NATO Grants PST.CLG976391
and CRG 972991. Other parts were done while S\'andor Fekete
was visiting MIT, with partial funding by DFG travel grant FE 407/7-1.

\end{document}

%% file: sandor.tex
\begin{picture}(0,0)%
\includegraphics{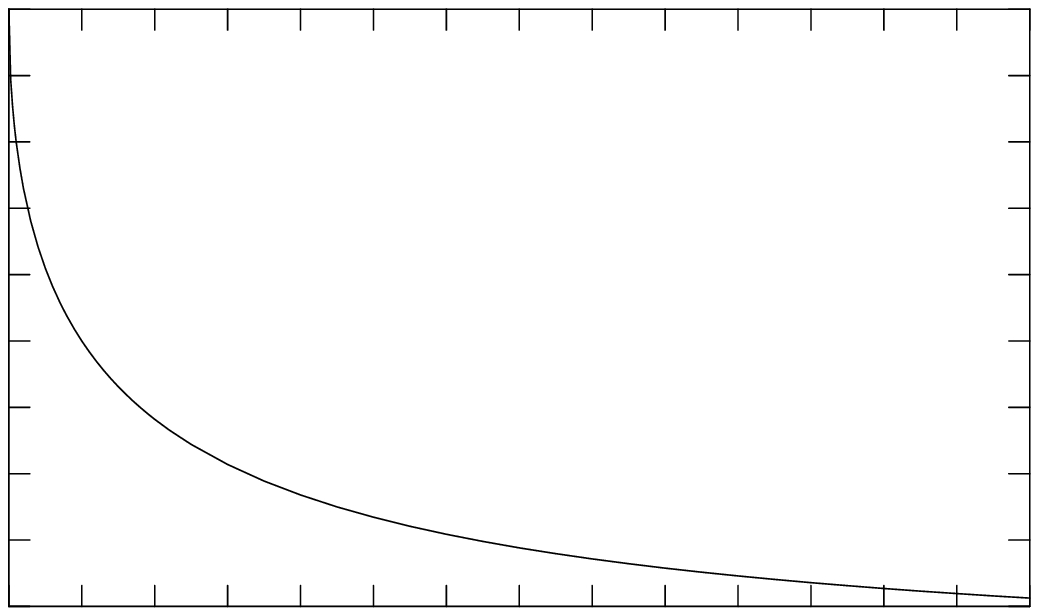}%
\end{picture}%
\setlength{\unitlength}{0.0200bp}%
\begin{picture}(18000,10800)(0,0)%
\put(2200,1650){\makebox(0,0)[r]{\strut{} 1.1}}%
\put(2200,2606){\makebox(0,0)[r]{\strut{} 1.2}}%
\put(2200,3561){\makebox(0,0)[r]{\strut{} 1.3}}%
\put(2200,4517){\makebox(0,0)[r]{\strut{} 1.4}}%
\put(2200,5472){\makebox(0,0)[r]{\strut{} 1.5}}%
\put(2200,6428){\makebox(0,0)[r]{\strut{} 1.6}}%
\put(2200,7383){\makebox(0,0)[r]{\strut{} 1.7}}%
\put(2200,8339){\makebox(0,0)[r]{\strut{} 1.8}}%
\put(2200,9294){\makebox(0,0)[r]{\strut{} 1.9}}%
\put(2200,10250){\makebox(0,0)[r]{\strut{} 2}}%
\put(2475,1100){\makebox(0,0){\strut{} 9}}%
\put(3525,1100){\makebox(0,0){\strut{} 10}}%
\put(4575,1100){\makebox(0,0){\strut{} 11}}%
\put(5625,1100){\makebox(0,0){\strut{} 12}}%
\put(6675,1100){\makebox(0,0){\strut{} 13}}%
\put(7725,1100){\makebox(0,0){\strut{} 14}}%
\put(8775,1100){\makebox(0,0){\strut{} 15}}%
\put(9825,1100){\makebox(0,0){\strut{} 16}}%
\put(10875,1100){\makebox(0,0){\strut{} 17}}%
\put(11925,1100){\makebox(0,0){\strut{} 18}}%
\put(12975,1100){\makebox(0,0){\strut{} 19}}%
\put(14025,1100){\makebox(0,0){\strut{} 20}}%
\put(15075,1100){\makebox(0,0){\strut{} 21}}%
\put(16125,1100){\makebox(0,0){\strut{} 22}}%
\put(17175,1100){\makebox(0,0){\strut{} 23}}%
\put(550,5950){\rotatebox{90}{\makebox(0,0){\strut{}Additive term over turn cost ($B/d$)}}}%
\put(9825,275){\makebox(0,0){\strut{}Competitive ratio ($c$)}}%
\end{picture}%
 